\documentstyle[aaspp4,flushrt]{article}
\input epsf.sty
\let\simlt=\lesssim
\let\simgt=\gtrsim
\newcommand{\fnu}{f_\nu}
\newcommand{\fb}{f_b}
\newcommand{\fc}{f_c}
\newcommand{\fcb}{f_{cb}}
\newcommand{\fnub}{f_{\nu b}}
\newcommand{\Td}{T_{\rm master}}
\renewcommand{\tilde}{\widetilde}
\begin{document}

\title{Small Scale Perturbations in a General MDM Cosmology}

\author{Wayne Hu\footnote{Alfred P. Sloan Fellow} and Daniel J.\ Eisenstein}
\affil{Institute for Advanced Study, Princeton, NJ 08540}

\begin{abstract}
For a
universe with massive neutrinos, cold dark matter, and baryons,
we solve the linear perturbation equations {\it analytically} 
in the small-scale limit 
and find agreement with numerical codes at the
$1-2\%$ level. The inclusion of baryons, a cosmological
constant, or spatial curvature reduces the small-scale power and
tightens limits on the neutrino density from observations of high
redshift objects.  Using the asymptotic solution, we investigate 
neutrino infall into potential wells and show that it can be 
described on {\it all} scales by a growth function that depends on 
time, wavenumber, and cosmological
parameters.  The growth function may be used to scale the 
present-day transfer functions back in redshift.
This allows us to construct 
the time-dependent transfer function for each species
from a single master function that is independent of
time, cosmological constant, and curvature.
\end{abstract}

\section{Introduction}

The mixed-dark-matter (MDM) scenario for structure formation involves a
hot component of massive neutrinos as well as the usual cold and
baryonic dark matter components.  In this case, even calculations in
linear perturbation theory are non-trivial due to the time-dependent
energy-momentum relation and non-vanishing angular moments of the
neutrino distribution.  Perturbations no longer grow uniformly with
time independent of scale.  Specifically, the growth of fluctuations is
suppressed below the time-dependent free-streaming scale of the
neutrinos due to collisionless damping.  
Numerical calculations with state-of-the-art Boltzmann
codes (e.g.\ \cite{Ma95}\ 1995; \cite{Dod96}\ 1996; \cite{Sel96}\ 1996) 
still require a fair amount of time to
solve the evolution equations on small spatial scales.  Moreover, the
additional parameter represented by the neutrino mass $m_\nu$
makes an exhaustive search of parameter space more difficult and has
led most workers to date to fix parameters such as the baryon density
(see e.g.\ \cite{Ma96}\ 1996).  For these reasons, we consider here an
analytic treatment of small-scale perturbation theory in MDM
cosmologies.

The inclusion of baryonic dark matter further complicates the
dynamics.  Recent measurements of high-redshift deuterium abundances
(\cite{Tyt96}\ 1996; but see \cite{Rug96}\ 1996) and new theoretical
interpretations of the Lyman--$\alpha$ forest (\cite{Wei97}\ 1997, and
references therein) suggest a value of the baryon density $\Omega_b$
greater than the fiducial nucleosynthesis value of $0.0125h^{-2}$
(\cite{Wal91}\ 1991).  Baryons suppress fluctuations on small scales
because, prior to recombination, photon pressure from the cosmic
microwave background supports them against collapse.  Hu \& Sugiyama
(1996, hereafter \cite{Hu96}) developed a formalism to account for this
effect and solve the evolution equations exactly on small scales.  The key aspect
to the treatment is the ability to ignore completely the role of
baryons as a gravitational source for enhancing CDM fluctuations.  

Massive neutrinos act in a manner similar to the pre-recombination baryons.
On scales smaller than their free-streaming length, the neutrinos are
smoothly distributed and hence do not contribute to the growth of
perturbations.  Here we generalize the techniques of \cite{Hu96}
to include the hot component, thereby allowing us to solve analytically 
for the transfer function on the smallest scales.  We then consider 
how to describe the end of free streaming and the 
resulting infall of neutrinos into the existing
potential wells.  This allows us to collapse all of the
late-time neutrino effects and base the transfer function 
on a single time-independent function of scale.

In an MDM cosmology with realistic baryon content,
the amplitude of small-scale fluctuations is important due to growing evidence
of early structure formation from high-redshift observations.  
The model has difficulty in explaining observations of 
galaxies at redshift $z \sim 3$ (\cite{Ste96}\ 1996; \cite{Mo96}\ 1996)
as well as damped Lyman-$\alpha$ systems at at comparable 
redshift (\cite{Mo94}\ 1994; \cite{Kau94}\ 1994; \cite{Kly95}\ 1995;
\cite{Ma97}\ 1997).
Baryons only exacerbate this problem and tighten the upper limit
on $\Omega_\nu$.  
Indeed, they yield a stronger effect for MDM as compared to CDM
cosmologies because $\Omega_b/(\Omega_0-\Omega_\nu)$ rather
than $\Omega_b/\Omega_0$ enters into the fluctuation amplitude.
Similarly, the growth rate of fluctuations is
determined by $\Omega_\nu/\Omega_0$ such that a given 
$\Omega_\nu$ causes more suppression in a low-density universe.
Our results here should therefore aid in the investigation
of the parameter space left available to MDM cosmologies. 

The outline of this paper is as follows.  After establishing the
notation in \S \ref{sec:notation}, we present in \S \ref{sec:small} 
the small-scale solutions
of the perturbation equations derived in the Appendix.  
We use these solutions in \S \ref{sec:infall} to
study the behavior of neutrino infall and to find analytic
approximations thereof.  From these results, we construct in \S
\ref{sec:transfer} the transfer functions in time and wavenumber for
the cold dark matter and total density perturbations and find 
agreement at the percent level 
with analytic results in the small-scale limit.  In \S
\ref{sec:lambda}, we show how these results may be scaled to
cosmologies with a cosmological constant or spatial curvature.

\section{Notation}
\label{sec:notation}

We begin by establishing the notation used throughout.
The density of the $i$th particle species ($i=c$, cold dark matter;
$b$, baryonic dark matter; $\nu$, massive neutrinos)
today in units of the
critical density is denoted $\Omega_i$, whereas the fraction of
the total matter density today $\Omega_0 = \sum_i \Omega_i$ is denoted
$f_i = \Omega_i/\Omega_0$.
As short-hand, we employ for example $i=cb$ to denote $\fcb =\fc + \fb$.
Note that $\fc+\fb+\fnu=1$.
Density perturbations are expressed as
$\delta\rho_i/\rho_i = \delta_i$,  
where the hybrid combinations 
are density weighted 
(e.g.~$\delta_{cb} = \fc \delta_c + \fb \delta_b$).  
The CMB temperature is given by $T_{\rm CMB} =
2.7 \Theta_{2.7}{\rm\,K}$; the best determination to date is
$2.728 \pm 0.004 {\rm \,K}$
(\cite{Fix96}\ 1996; 95\% confidence interval) 
at which it is fixed for most of our expressions.
Finally, as usual the Hubble constant is written as 
$H_0 = 100h{\rm\,km\,s^{-1}\,Mpc^{-1}}$.

Time is parameterized as
$y=(1+z_{\rm eq})/(1+z)$,
where
\begin{equation}\label{eq:zeq}
z_{\rm eq} = 2.50 \times 10^4 \Omega_0 h^2 \Theta_{2.7}^{-4}
\end{equation}
is the redshift of matter-radiation equality.  The second 
important epoch is when the baryons are released from 
the Compton drag of the photons near recombination,
i.e.~$y_d=y(z_d)$ where (\cite{Hu96}; \cite{Eis97a}\ 1997a) 
\begin{eqnarray}\label{eq:zdrag}
z_d & = & 1291 {(\Omega_0 h^2)^{0.251}
\over 1 + 0.659 (\Omega_0 h^2)^{0.828} }
[1 + b_1 (\Omega_b h^2)^{b_2}]\,, \\
b_1 & = & 0.313 (\Omega_0 h^2)^{-0.419} [1 + 0.607
(\Omega_0 h^2)^{0.674} ]\,, \nonumber\\
b_2 & = & 0.238 (\Omega_0 h^2)^{0.223}\,.\nonumber
\end{eqnarray}
After this epoch, baryons fall into the potential wells provided
by the cold dark matter and participate in gravitational collapse.

We often label the comoving wavenumber $k$
relative to the scale that
crosses the horizon at matter-radiation equality, thus defining the quantity
\begin{equation}
q = {k \over {\rm Mpc}^{-1}} \Theta_{2.7}^2 (\Omega_0 h^2)^{-1}
  = {k \over 19.0} (\Omega_0 H_0^2)^{-1/2} (1+z_{\rm eq})^{-1/2}\,.
\label{eq:q}
\end{equation}
The small-scale limit is defined as $q \gg 1$. 
In the next section (\S {\ref{sec:small}), we place an 
additional restriction that the momentum of the neutrinos 
keep them
out of the perturbations formed by the heavier species.  
Such scales are below the {\it free-streaming} scale, which itself
shrinks with
time [c.f.\ eq.\ (\ref{eq:yfs})]. We show in 
\S\ref{sec:infall} how to account for neutrino infall.

We often encounter functions of wavenumber or time that depend 
additionally on cosmological parameters; we denote these as
e.g.
$F(y,q;\fnu\ldots)$. 
Where the cosmological parameter dependence is not being emphasized,
we often drop the parameters after the semicolon, e.g.\ 
$F(y,q)
=F(y,q;\fnu\ldots).$

\section{Small Scale Solution}
\label{sec:small}

Below the free-streaming scale of the neutrinos (see \S \ref{sec:infall}) 
and sound horizon of the baryons at recombination, the equations of
motion for matter density fluctuations may be solved analytically in a
matter $+$ radiation universe using the techniques of \cite{Hu96}.
The key approximation is that on sufficiently small scales the neutrinos
move too quickly to trace the perturbations in the CDM and baryons.
In this case, the neutrinos contribute no gravitational sources to the
evolution equations of the other species, thereby slowing the growth
of fluctuations.  The baryons have a similar behavior prior to the
drag epoch; in the Appendix we describe how to include both effects.
The result is that density perturbations grow as
\begin{equation}
\delta_{cb}(y,q;\fnu,\fb,y_d) = D_{cb}(y;\fnu)
\delta_d(q;\fnu,\fb,y_d) \,.
\label{eqn:deltacb}
\end{equation}
Equation (\ref{eqn:deltacb}) states that the 
density perturbation today is the product of
a growth function $D_{c b}$ that depends on the neutrino fraction and
the amplitude $\delta_d$ of fluctuations entering the growing mode 
at the Compton drag epoch $y_d$.  

The quantity
\begin{equation}
p_i(f_i) = {1\over4}\left[5-\sqrt{1+24f_i}\right] \ge 0 
\end{equation}
determines the reduction from a linear growth rate $\delta \propto
y$, where $f_i$ is the fractional density
in gravitationally clustering components: $i=c$ and
$i=cb$ 
before and after the drag epoch respectively.
Hence the growth factor is given by
\cite{Bon80} (1980)
\begin{equation}
D_{cb}(y;\fnu) = y^{1 - p_{cb}},
\end{equation}
where we take $\Omega_0=1$; we generalize to $\Omega_0\ne1$ in
\S \ref{sec:lambda}.

The amplitude of the fluctuation entering the growing mode at the drag epoch
is 
\begin{equation}
\delta_d(q;\fnu,\fb,y_d)= 9.50 M_d(q) \Phi(0,q) \, ,
\label{eqn:deltad}
\end{equation}
where $\Phi(0,q)$ is the initial amplitude of the potential perturbation.
The quantity $M_d$ expresses the matching condition between
the growing mode $y^{1 - p_c}$ before the drag epoch and
$y^{1 - p_{cb}}$ after the drag epoch, as well as the matching required
to describe the onset of matter domination.  We find
\begin{eqnarray}
M_d(q;\fnu,\fb,y_d) &=&
{\fc \over \fcb} {5-2(p_{c}+p_{cb})\over 5-4p_{cb}} (1+y_d)^{p_{cb}-p_{c}}
\nonumber\\
&& \times \left[ 1 + {p_{c}-p_{cb} \over 2} \left(1+{1\over{(3-4p_{c})
(7-4p_{cb})}} \right) (1+y_d)^{-1} \right]\, A_1(q)\,.
\label{eqn:Md}
\end{eqnarray}
where\footnote{We assume here the number of massive neutrinos $N_\nu=1$;
see equation~(\ref{eqn:correction}) for the general case.}
\begin{equation}
A_1(q;\fnub,\fnu) = { 1 - 0.553 \fnub + 0.126 \fnub^3 \over 
	1 - 0.193  \fnu^{1/2} + 0.169 \fnu } \ln ({1.84 \beta_c q})\,,
\end{equation}
with
\begin{equation}
\beta_c = [1- 0.949 \fnub]^{-1}\,.
\end{equation}
Equation (\ref{eqn:Md}) results from a series expansion in
$(1+y_d)^{-1}$ of the analytic solution and accounts for small
deviations from the power-law growing-mode behavior due to radiation.
The expansion is only accurate for $\fb+\fnu\lesssim0.6$ and $y_d\gtrsim1$,
but the form in the Appendix is general.
Notice that $M_d \rightarrow A_1$ as $\fb \rightarrow 0$ and that the
term in brackets introduces only a small correction since $y_d \gtrsim
1$ in cases of interest.

\begin{figure}[bt]
\centerline{\epsfxsize=3.5truein 
\epsffile{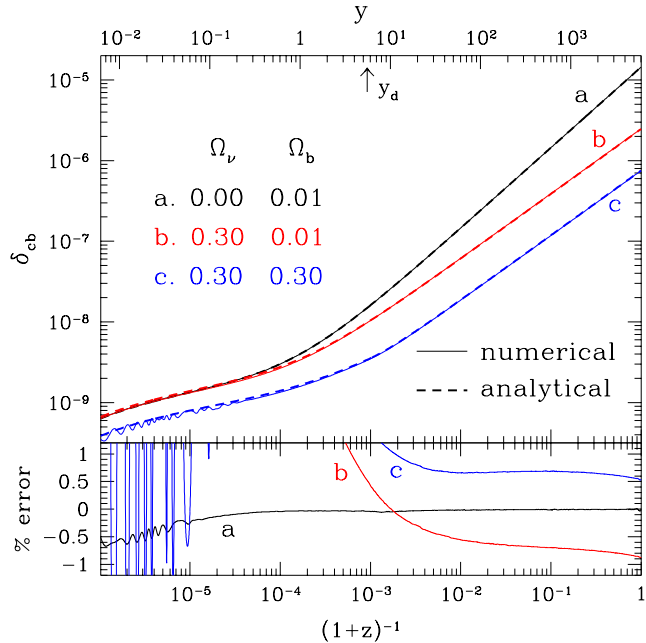}}
\caption{Growth suppression from the neutrinos and baryons.  The
addition of neutrinos slows the growth of CDM $+$ baryon fluctuations
after matter-radiation equality $y>1$ (upper panel, curve $b$ compared with
curve $b$).
Increasing the baryon fraction further suppresses fluctuations by a
time-independent factor at $y>y_d$ due to growth suppression in the CDM
fluctuations for $1<y<y_d$ and the lack of baryon fluctuations in the
weighted density perturbations (upper panel, curve $c$).  The
analytic expressions (dashed lines) agree with numerical results
(solid) for $\Omega_0=1$, $h=0.5$ and $q=160$ at the $1\%$ level except
for $5\%$ discrepancies at $y \le 1$.  }
\label{fig:growth}
\end{figure}

In Fig.~\ref{fig:growth}, we display an example of the time evolution
of a mode given by the analytic solution (including here the decaying
mode given in the Appendix, important for $y \simlt y_d$) compared with
numerical solutions.  The $5\%$ offset at early times is due to changes
in the expansion rate as the neutrinos become non-relativistic.  The
oscillatory errors arise from neglect of baryon acoustic oscillations;
these Silk damp away well before the drag epoch for these scales.  
The main effect
of the neutrinos is to slow the growth of the CDM and baryons after
equality $(y > 1)$ because they represent a smooth gravitationally-stable 
component on these scales.  
Similarly,
since the baryons have no fluctuations on
these scales until they fall into CDM potential wells after the drag epoch
$y_d$, they reduce the growth rate 
between equality and the drag epoch.
Furthermore, they reduce the
net fluctuation $\delta_{cb}$ 
 as $\fc / \fcb$ leading to the offset between the curves
in Fig.~\ref{fig:growth}.

\section{Neutrino Infall}
\label{sec:infall}

Eventually, the neutrinos fall into the CDM potential wells, breaking
the approximation of the last section.  The neutrino thermal velocity
decays with the expansion of the universe as $v_\nu \propto (a
m_\nu)^{-1}$; infall occurs when their velocity slows sufficiently as
to allow clustering by the Jeans criteria (\cite{Bon83}\ 1983;
\cite{Ma96}\ 1996):
\begin{eqnarray}
k_{\rm fs} &=& 4\pi G\rho a^2/v_\nu^2\propto(1+z)^{-1/2}(\Omega_0 h^2)^{3/2} 
(\fnu /N_\nu)\,, \nonumber\\
q_{\rm fs} &\propto& y^{1/2} (\fnu/N_\nu)  \,.
\label{eqn:kfs}
\end{eqnarray}
Recall that $k$ and $q$ are related by equation ({\ref{eq:q}).
Here
$N_\nu$ is the number of massive neutrino species, assumed to
be degenerate in mass.  For simplicity, we restrict our examples
to $N_\nu=1$ throughout, but we have verified 
that the infall description 
is valid for $N_\nu\ne1$ at the $1-2\%$ level
in the range $0 \le z \le 25$.\footnote{Effects
at redshifts approaching the epoch at which the neutrinos become
non-relativistic are not accounted for by this approximation. For $N_\nu \ne 1$,
the growth function of equation (\ref{eqn:infallgrowth}) allows scaling
at low redshifts but this does not imply $\delta_d$ or $\Td$ is independent of
$N_\nu$ around the maximal infall scale.  
We treat these effects in \cite{Eis97b} (1997b).}

On scales $q
\ll q_{\rm fs}$, the neutrinos will follow the cold dark matter.  By
acquiring density perturbations, they enhance the CDM$+$baryon
potential wells and drive the growth rate back up to $y$.  The CDM$+$
baryon density fluctuations thus acquire a scale dependence to their
growth rate
\begin{equation}
D_{cb}(y,q;\fnu) 
= \left[1+\left(y\over1+y_{\rm fs}(q;\fnu)\right)^{0.7} 
\right]^{p_{cb}/0.7} y^{1-p_{cb}} \,,
\label{eqn:infallgrowth}
\end{equation}
where the coefficient $0.7$ represents a fit to the numerical evolution.  

The transition epoch $y_{\rm fs}$ incorporates two effects.  
The first is that from equation (\ref{eqn:kfs}) the characteristic
epoch for infall for a given wavenumber must scales as 
$y\propto(qN_\nu/\fnu)^2$.
The second is that the growing modes of the free-streaming epoch 
$\propto y^{1-p_{cb}}$ and the infall epoch $\propto y$
must be matched across the transition.
This matching condition may only depend on $\fnu$. 
We find that the total effect is well approximated by 
\begin{equation}\label{eq:yfs}
y_{\rm fs}(q;f_\nu) = 17.2 \fnu ( 1+ 0.488 \fnu^{-7/6} ) (q N_\nu/\fnu)^2\,, 
\end{equation}
In Fig.~\ref{fig:infall} we show the enhancement of density fluctuations due to
infall and compare $D_{cb}$ from equation~(\ref{eqn:infallgrowth})  
to numerical solutions for various $q$ (short-dashed lines). 

\begin{figure}[bt]
\centerline{\epsfxsize=3.5truein 
\epsffile{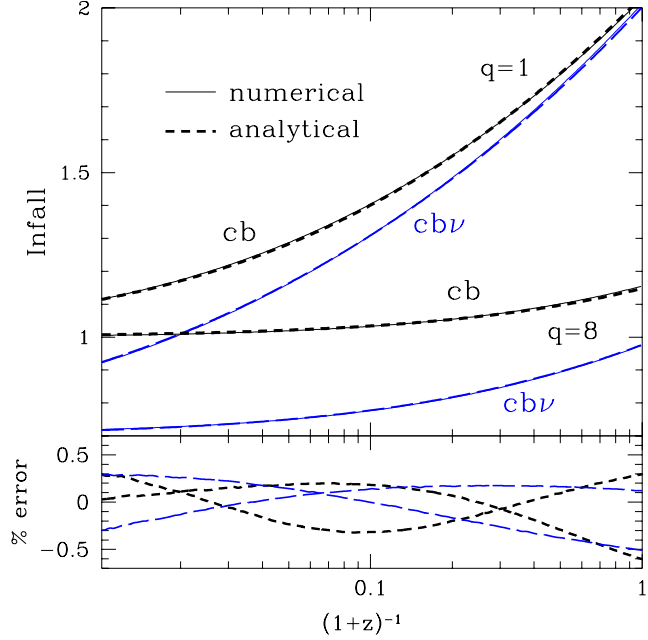}}
\caption{Analytic vs.\ numerical descriptions of neutrino infall for 
$\fnu=0.3$, $\fb=0.01$, $\Omega_0=1$, $h=0.5$. 
Upper panel: Growth functions compared to the fully free-streaming
time-dependence of $y^{1-p_{cb}}$;  
numerical CDM $+$ baryon fluctuations compared 
with growth function $D_{cb}$
({\it short-dashed lines}) 
numerical CDM $+$ baryon $+$ neutrino 
density-weighted fluctuations compared with $D_{cb\nu}$
({\it long-dashed lines}).  No infall is represented by a horizontal line
of amplitude 1 ($cb$) and $f_{cb}=0.69$ ($cb\nu$) here.
Lower panel: relative error.}
\label{fig:infall}
\end{figure}

Furthermore, the density-weighted fluctuation
\begin{equation}
\delta_{cb\nu} = \fcb\delta_{cb} +\fnu\delta_{\nu}
\end{equation} 
follows from the infall solution by noting that
it converges above the free-streaming scale to $\delta_{cb}$ and
below to $\fcb\delta_{cb}$.  Thus
\begin{equation}
\delta_{cb\nu}(y,q;\fnu,\fb,y_d) = 
D_{cb\nu}(y,q;\fnu) \delta_d(q;\fnu,\fb,y_d) \,,
\end{equation}
where
\begin{equation}
D_{cb\nu}(y,q;\fnu) = \left[ \fcb^{0.7/p_{cb}} 
	+ \left( { y \over 1 + 
	y_{\rm fs}(q;\fnu) }\right)^{0.7} \right]^{p_{cb}/0.7} 
	y^{1-p_{cb}} .
\end{equation}
In Fig.~\ref{fig:infall},
we show that this form produces a good fit to the numerical results 
(lower set, long-dashed lines). 

Finally, the horizon at the epoch when the neutrinos become
nonrelativistic sets the maximal free-streaming scale.  Beyond this
scale, the neutrinos are always in the infall regime and the evolution
of density fluctuations becomes independent of the neutrino fraction
\begin{equation}
\lim_{q\rightarrow 0} \delta_{cb}(y,q) 
	= {1 \over 137 q^2} y \Phi(0,q) \,.
\label{eqn:deltaq0}
\end{equation}
The appearance of $y/(1+y_{\rm fs})$ in equation (\ref{eqn:infallgrowth}) 
assures the proper time dependence for the evolution in the 
large-scale limit.  The growth functions are thus not subject to a small-scale
approximation and remain valid for all $q$.

\begin{figure}[bt]
\centerline{\epsfxsize=3.5truein 
\epsffile{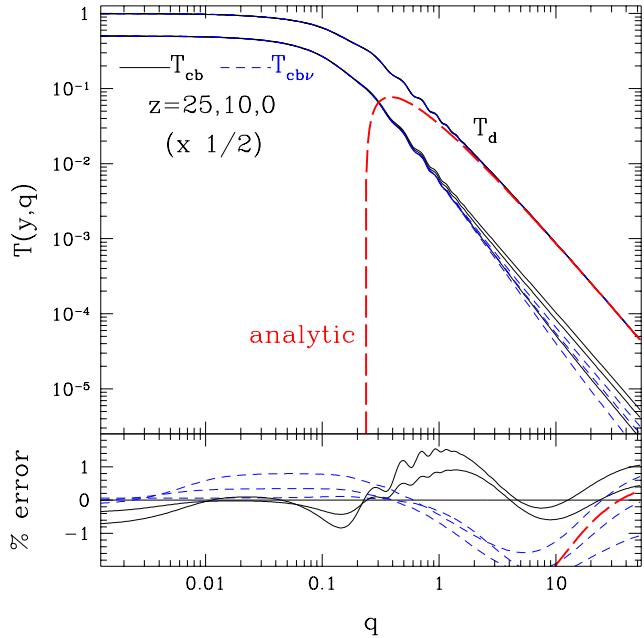}}
\caption{Demonstration of the existence of a time-independent 
master function.   The transfer functions for the CDM $+$ baryon 
($T_{cb}$) and CDM $+$ baryon $+$ neutrino ($T_{cb\nu}$) fluctuations
at 3 different redshifts are divided by the growth
factors $D_{cb}$ and $D_{cb\nu}$ respectively to obtain estimates of $\Td$.  
That the 6 curves superimposed in the upper
panel agree at the $1-2\%$ level (relative to $T_{cb}/D_{cb}$ at
$z=0$, as shown in the lower panel) establishes the existence of 
the master function and verifies the accuracy of the
growth functions.  The analytic prediction for $\Td$ (long-dashed line) 
converges to within $1\%$ of the numerical results at $q\gg 1$.  The model
here is $\Omega_0=1$, $h=0.5$, $\fnu=0.4$, $\fb=0.2$.
 }
\label{fig:transfer}
\end{figure}
\section{Transfer Functions}
\label{sec:transfer}

We are now in a position to evaluate the transfer function, defined as
\begin{equation}
T_i(y,q) = {\delta_i(y,q) \over \delta_i(y,0)} 
{\Phi(0,0) \over \Phi(0,q)}  \,,
\end{equation}
for the $i$th component of the matter.  For the CDM $+$ baryon  ($i=cb$)
and the CDM $+$ baryon $+$ neutrino ($i=cb\nu$) systems,
we obtain
\begin{eqnarray}
T_{cb}(y,q;\fnu,\fb,y_d)& = & y^{-1} D_{cb}(y,q; \fnu) 
	\Td(q;\fnu,\fb,y_d) \,,\nonumber\\
T_{cb\nu}(y,q;\fnu,\fb,y_d)& = & y^{-1} D_{cb\nu}(y,q;\fnu) 
	\Td(q;\fnu,\fb,y_d)  \,,
\label{eqn:Tcbn}
\end{eqnarray}
where $\Td$ has the small-scale limit of
\begin{equation}
\lim_{q \rightarrow \infty} \Td(q;\fnu,\fb,y_d)=
	{M_d(q) \over 14.4 q^2} \,,
\label{eqn:T_d}
\end{equation}
which follows from equations (\ref{eqn:deltad}) and
(\ref{eqn:deltaq0}).  It is important to note that relation
(\ref{eqn:Tcbn}) holds independently of the small-scale approximation:
namely that once the growth factors $y^{-1} D_i$ are removed, the
transfer functions depend only on a {\it single} time-independent
function of $q$ related to perturbations in the CDM component at the
drag epoch.  This simplification holds only for $y\gg y_d$; near the
drag epoch, the contribution of the decaying mode cannot be neglected.
In \cite{Eis97b} (1997b), we exploit the existence of this
master function to obtain the full time- and $q$-dependent transfer
functions based on fitting formulae for $\Td$.

We show a comparison between analytic and numerical results in Fig.~\ref{fig:transfer}.
The numerical $T_{cb}, T_{cb\nu}$ (lower curves, solid and dashed lines respectively) 
at 3 different redshifts are plotted here.  The upper curve represents
the master function $\Td$ obtained through inverting the growth
factor.  Note that 6
different estimates of $\Td$ obtained 
from $T_{cb}$ and $T_{cb\nu}$ are superimposed and
agree at the $\sim 1-2\%$ level.
Also shown is the small-scale prediction of equation~(\ref{eqn:T_d}) which converges rapidly for
$q \simgt 1$.

Finally, note that the neutrino transfer function is implicitly defined as
\begin{equation}
T_\nu = f_\nu^{-1} (T_{cb\nu} - f_{cb} T_{cb})\,,
\end{equation}
and its construction from the growth functions and $\Td$ yields {\it density-weighted}
errors on the same order as the other transfer functions, i.e. $(1-2\%)\times
\,\delta_{cb\nu}/f_\nu\delta_\nu$.

\section{Low Density Models}
\label{sec:lambda}

The formulae presented thus far are valid for $\Omega_0=1$ 
cosmologies.  We have however shown that the transfer functions today
can be expressed as the products of a growth function and a master function 
related to fluctuations at the drag epoch.  At this earlier time,
the cosmological constant and curvature have negligible effects
on the dynamics.
This implies that a simple modification of the growth function
at late times to account for $\Omega_0 \ne 1$ effects will 
suffice for a complete description.

Let us recall that on the largest scales, where neutrino free
streaming and radiation pressure gradients are negligible, fluctuations 
grow as (\cite{Hea77}\ 1977)
\begin{equation}
D(y;\Omega_0,\Omega_\Lambda) 
= {5 \over 2} g(y) \int^y {dx \over x^3 g(x)^3}\,,
\label{eqn:Dgrowth}
\end{equation}
where
\begin{equation}
g^2(y) = y^{-3} + y^{-2} y_0^{-1} (1-\Omega_0-\Omega_\Lambda)/\Omega_0
	+ y_0^{-3} \Omega_\Lambda/\Omega_0\,,
\end{equation}
with $y_0=(1+z_{\rm eq})$.
Analytic forms for 
the $\Omega_\Lambda=0$ and $\Omega_0+\Omega_\Lambda=1$ cases
are given in \cite{Gro75} (1975) and \cite{Bil92} (1992) respectively.
The normalization of the growth rate  
has been chosen so that $D=y$ at early times.  Moreover,
equation (\ref{eqn:Dgrowth}) states that after matter ceases to
dominate the expansion rate, fluctuation growth halts.  

By matching these asymptotic limits, we can approximate the
growth function in the presence of neutrinos by the replacement
of $y$ with $D$, i.e.
\begin{eqnarray}
D_{cb}(y,q;\fnu,\Omega_0,\Omega_\Lambda)
	&=&\left[1+\left(D(y)\over1+y_{\rm fs}(q)\right)^{0.7}\right]
	^{p_{cb}/0.7} D^{1-p_{cb}}(y)  \,, \\
D_{cb\nu}(y,q;\fnu,\Omega_0,\Omega_\Lambda) 
	&=& \left[\fcb^{0.7/p_{cb}} 
	+\left(D(y)\over1+y_{\rm fs}(q)\right)^{0.7}\right]^{p_{cb}/0.7} 
	D^{1-p_{cb}}(y)\,,
\label{eqn:growthlambda}
\end{eqnarray}
where of course $D(y)= D(y;\Omega_0,\Omega_\Lambda)$ implicitly.
Likewise, the transfer functions become
\begin{eqnarray}
T_{cb}(y,q)& = & D^{-1}(y) D_{cb}(y,q) 
	\Td(q) \,,\nonumber\\
T_{cb\nu}(y,q)& = & D^{-1}(y) D_{cb\nu}(y,q) 
	\Td(q)  \,.
\label{eqn:TcbnD}
\end{eqnarray}

We show an example in Fig.~\ref{fig:lambda}.  The model has been chosen
to have $\Omega_0=0.25$, $\Omega_\Lambda=0.75$ with 
the same $\Omega_0 h^2$, $\fb$ and $\fnu$ as Fig.~\ref{fig:transfer}
and hence the same $\Td$.  The invariance of the master function is
demonstrated by overplotting estimates from $T_{cb}$ and $T_{cb\nu}$ in
this model and from $T_{cb}$ of Fig.~\ref{fig:transfer}.  The lower
panel shows the fractional difference of the former with respect to the
latter.  

In principle, the fact that infall freezes out at a later redshift
in a $\Lambda$ vs. open cosmology of the same $\Omega_0$ allows
one to distinguish between these two alternatives by the shape of the
transfer function at a single redshift alone.  Massive neutrinos
therefore break the {\it shape degeneracy} of the transfer function in 
low-density universes.  
However, since this effect is only significant if $\Omega_0\ll 1$, the
more important fact is that growth in the matter-dominated
epoch depends on
$(1-\Omega_\nu/\Omega_0)$ and hence even a relatively
small density in neutrinos can make a dramatic effect on the
growth rates in a low-density universe.   Fig.~\ref{fig:lambda} demonstrates
that $\Omega_\nu=0.1$ in an $\Omega_0=0.25$ universe gives the same magnitude
effects as $\Omega_\nu=0.4$ in an $\Omega_0=1$ universe.  Upper limits 
on $\Omega_\nu$ from small-scale fluctuations thus tighten for low
density universes (\cite{Pri95}\ 1995).

\begin{figure}[bt]
\centerline{\epsfxsize=3.5truein 
\epsffile{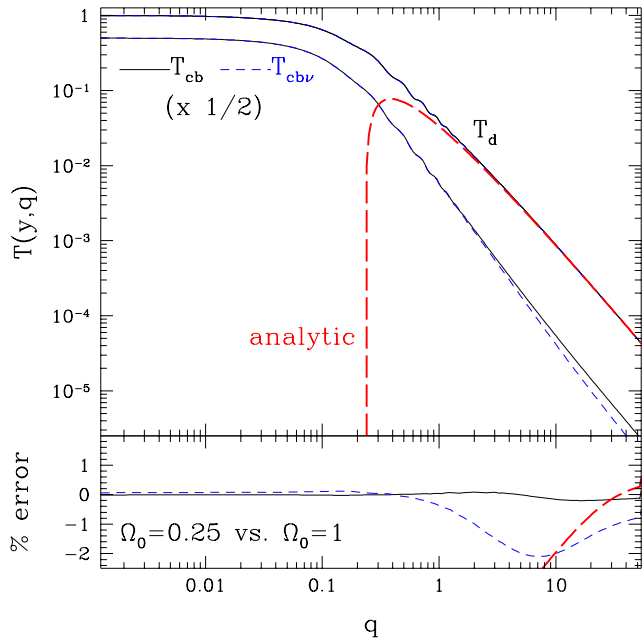}}
\caption{Demonstration of the invariance of the master 
function ($\Td$) under changes in
the cosmological constant $\Lambda$. The model has the same 
$\Omega_0 h^2$, $\fb$ and $\fnu$ as Fig.~\protect{\ref{fig:transfer}}
at $z=0$ but with $\Omega_0=0.25$.
Division of $T_{cb}$ and $T_{cb\nu}$ by the growth functions
returns the master function $\Td$ over which 
the estimate of Fig.~\protect{\ref{fig:transfer}} and the small-scale analytic solution are
plotted (top panel).   The errors are plotted relative to $\Td$ from the
$T_{cb}$, $z=0$ estimate of Fig.~\protect{\ref{fig:transfer}} and show agreement 
at the $1-2\%$ level (bottom panel).  
 }
\label{fig:lambda}
\end{figure}

\section{Discussion}
\label{sec:discussion}

We have presented small-scale solutions to the linear evolution of
perturbations in a cold $+$ hot $+$ baryonic dark matter cosmology.
They converge to within $1\%$ of the numerical solutions for $q \gg 1$
and $q \gg q_{\rm fs}$  and are expressed in terms of hypergeometric
functions in the Appendix.  We have also given simplified forms in the
main text, employing only algebraic functions, which are valid for
$\Omega_b + \Omega_\nu \simlt 0.6\Omega_0$ and $\Omega_0 h^2 \simgt 0.1$.
Baryons play a significant role in MDM cosmologies since 
dark-matter perturbations
are density weighted and depend on $\Omega_b/(\Omega_0-\Omega_\nu)$
not $\Omega_b/\Omega_0$.  Likewise neutrinos in a low-density 
universe yield enhanced effects since growth
rates depend on $\Omega_\nu/\Omega_0$. 

By comparing analytic and numerical results at  
$q \simlt q_{\rm fs}$, we have isolated the effects of neutrino infall 
and described them to an accuracy of $1-2\%$ by time- and
wavenumber-dependent growth factors for the CDM and total matter.  
The freeze-out of infall at 
late times when $\Lambda$ or curvature come to dominate can similarly 
be taken into account.  
We have shown that these growth factors are valid
beyond the small-scale approximation.

The full time
and wavenumber dependent transfer functions for the CDM and total
matter can thus be described as a product of these growth factors
and a single function of wavenumber related to the amplitude of 
fluctuations at the Compton drag epoch that in turn depends on 
$\Omega_0 h^2$, $\Omega_\nu/\Omega_0$, $\Omega_b/\Omega_0$, and
the number of degenerate neutrino species.  We leave a quantitative
description of this master function and implications for
constraints on $\Omega_\nu$ to a companion piece (\cite{Eis97b}\ 1997b).

\noindent {\it  Acknowledgments:} W.H.\ and D.J.E.\ are supported by NSF
PHY-9513835.  W.H.\ was additionally supported by the W.M.\ Keck Foundation.
Numerical results were extracted from the CMBfast package of 
\cite{Sel96} (1996)~v. 2.3.
 
\appendix

\section{Derivation of the Small Scale Solution}

Following the analytic approach of \cite{Hu96}, one can solve the
equation of motion for the CDM on small scales where the gravitational
effects of the MDM can be neglected.  The idea is to separate the
evolution before and after the drag epoch.  Before the drag epoch, the
gravitational effects of the baryons can be ignored below the
sound horizon as they are pressure supported by the photons.  The
equation of motion then becomes
\begin{equation}
\ddot\delta_c + {\dot a \over a}\dot\delta_c = 
	{3 \over 2 a} \fc \Omega_0 H_0^2 \delta_c \, ,
\label{eq:deltaceta}
\end{equation}
where overdots indicate derivatives with respect to conformal time $\eta$.
Unfortunately, the complicated equation of state of massive neutrinos
prevents the time evolution of the scale factor $a(\eta)$ from
being written down in closed form.  However, we know that the
neutrinos behave as radiation at early times when $T_\nu \gg m_\nu$ and
as matter at late times when $T_\nu \ll m_\nu$.  These limits are
identical to those found if one considered the neutrinos to be massless
and added their mass to that of the non-relativistic matter.  
It is therefore a
reasonable first approximation to leave the background evolution
unmodified by the replacement of a portion of the non-relativistic
matter with massive neutrinos, i.e.
\begin{equation}
\eta = 2(\Omega_0 H_0^2)^{-1/2} (1+z_{\rm eq})^{-1/2} \left( \sqrt{ 
1+ y} -1 \right) \,,
\end{equation}
where $y=(1+z_{\rm eq})/(1+z)$.  Here, $z_{\rm eq}$ [eq.\ (\ref{eq:zeq})]
assumes three {\it massless} neutrino species with the usual thermal history.  
We neglect cosmological constant and
curvature effects here (see \S \ref{sec:lambda}).  
This form otherwise
errs only between $z_{\rm eq}$ and the
epoch at which the neutrinos become non-relativistic.
Our approach will be to use this approximation to
solve the small-scale limit analytically and then correct for a $\simlt
5\%$ modification due to changes in the expansion rate.

With this approximation, equation (\ref{eq:deltaceta}) can be rewritten in
terms of $y \equiv (1+z_{\rm eq})/(1+z)$ as
\begin{equation}
{d^2 \over dy^2}\delta_c
+{(2+3y) \over 2y(1+y)}{d \over dy}\delta_c =
{3 \over 2 y(1+y)} \fc \delta_c \, .
\label{eq:deltacy}
\end{equation}
As shown in \cite{Hu96}, the general solution to this equation is given
through Gauss's hypergeometric function $F$ (also written ${}_2 F_1$) 
\begin{equation}
U_i =  (1 + y)^{-\alpha_i} F(\alpha_i,\alpha_i+{1 \over 2},
2\alpha_i + {1 \over 2} ; {1 \over 1 + y}) \, ,
\label{eq:hypersolution}
\end{equation}
where $i=1, 2$ and
\begin{equation}
\alpha_i = {1 \mp \sqrt{1 +24 \fc} \over 4} \, ,
\end{equation}
with $-$ and $+$ for $i=1$ and $2$, respectively.  It is useful to note
that $\lim_{y \rightarrow \infty} U_i = y^{-\alpha_i}$ and thus the two
solutions represent the growing and decaying mode for $i=1,2$
respectively.  Clearly, $\alpha_1+\alpha_2=1/2$.

We obtain the amplitude of the growing and decaying mode 
by matching onto the
$y \ll 1$ solution of \cite{Hu96} eq.~(B12)\footnote{The numerical factors here
reflect the kick a perturbation gets at horizon crossing deep in the radiation-dominated 
epoch.  They are calibrated to agree at the 1\% level with
CMBfast v.~2.3 (high precision version) and represent a 1-2\% shift
versus the calibration of \cite{Hu95}\ (1995) based on \cite{Sug95}\ (1995).
Our calibration also matches the code of M. White
(\cite{Hu95}\ 1995) at the 1\% level.} 
\begin{equation}
\delta_c = 9.50 \ln ( 9.24 q y )\Phi(0,q) \,, \qquad y \ll 1 \,,
\end{equation}
to find
\begin{equation}
\delta_c(y,q) = 9.50 [A_1(q) U_1(y) + A_2(q) U_2(y)] \Phi(0,q) \qquad y < y_d \,,
\label{eq:deltacyltyd}
\end{equation}
where the matching coefficients are
\begin{equation}
A_1(q)  =   {\Gamma(\alpha_1)\Gamma(\alpha_1+1/2)\over \Gamma(2\alpha_1+1/2)
 2\pi \cot (2\pi \alpha_1) } \left[ \ln(9.04 q)
+2\psi(1) - 2\psi(1-2\alpha_1) -2 \ln 2 \right]
\, , 
\label{eqn:A1}
\end{equation}
with $\psi(x)=\Gamma'(x)/\Gamma(x)$.  The expression for 
$A_2$ follows from equation~(\ref{eqn:A1}) with the
replacement $1 \rightarrow 2$ in the subscripts.

Equation (\ref{eq:deltacyltyd}) takes the evolution up to the drag epoch. 
After the drag epoch, the baryons are released from the photons and behave as 
cold dark matter.  The equation of motion for the combined CDM+baryon system
is the same as equation~(\ref{eq:deltacy}) but with the replacements
\begin{eqnarray}
\delta_c  &\rightarrow & {\fc \over \fcb} \delta_c + {\fb \over \fcb} \delta_b
	\equiv \delta_{cb} \,,\nonumber\\
\fc & \rightarrow & \fcb\,,
\end{eqnarray}
and therefore has the same solutions as given by equation 
(\ref{eq:hypersolution}) with
arguments
\begin{equation}
\tilde \alpha_i = {1 \pm \sqrt{1 +24\fcb} \over 4} \, .
\end{equation}

The full solution for $\delta_{cb}$ is now found by matching 
the solution of equation (\ref{eq:deltacyltyd}) onto the new growing and decaying modes.
Since the decaying mode becomes unimportant for $y \gg y_d$,
\begin{eqnarray}
\delta_{cb}(y,q) =  9.50 M_d(q) 
		\tilde U_1(y)\Phi(0,q) \,,
\label{eq:deltacygtyd}
\end{eqnarray}
where
\begin{equation}
M_d(q)= {\fc \over \fcb}\left({ \tilde U_2' [A_1 U_1 + A_2 U_2] 
			- \tilde U_2 [A_1 U_1' + A_2 U_2'] 
		\over \tilde U_1 \tilde U_2' - \tilde U_1' \tilde U_2 } \right) \Bigg|_{y=y_d}\,,
\end{equation}
gives the matching condition.

Furthermore,  the baryons fall into CDM potential wells at the drag epoch
and subsequently follow them so that
\begin{equation}
\delta_{cb} = \delta_c = \delta_b\,, \qquad y \gg y_d\,.
\end{equation}

For high $\Omega_0 h^2 \simgt 0.1$ (or $h \simgt 0.3$), $y_d \gg 1$ so that
the decaying-mode contribution to equation~(\ref{eq:deltacygtyd}) coming through
$U_2$ is negligible.  Furthermore the growing mode can be expanded in powers
of $(1+y_d)^{-1}$ as 
\begin{eqnarray}
M_d(q) &\approx& {\fc \over \fcb}
{2 (\tilde\alpha_1 + \alpha_1) - 1
\over 4 \tilde \alpha_1 - 1 } 
	(1+ y_d)^{\tilde\alpha_1 - \alpha_1} \nonumber\\
&& \times\left[ 1+ (1+y_d)^{-1}
{\alpha_1 - \tilde\alpha_1 \over 2} \left( 1 +  {1 \over {(4\alpha_1 + 1)
(4\tilde\alpha_1-3)}} \right)
	\right] A_1(q) \,.
\end{eqnarray}
The approximation holds for $\fc \simgt 0.2$.  Note that the apparent divergence
at $\fc = 0.125$ or $\alpha_1=-1/4$ disappears once the decaying mode is properly
included.

The quantity $A_1$ can either be evaluated from equation~(\ref{eqn:A1}) or approximated
as
\begin{equation}
A_1(q) = \left[ { 1 - 0.553 \fnub + 0.126 \fnub^3 } \right] \ln (1.84 \beta_c q) \,,
\end{equation}
with
\begin{equation}
\beta_c \equiv \exp[-2\psi(2\alpha_2)-2\psi(3)] 
	\approx [1 - 0.949\fnub]^{-1} \,,
\end{equation}
which is valid at the $1\%$ level for $\fnub \simlt 2/3$.  Finally, we
correct for the change in the background expansion rate by comparing
equation~(\ref{eq:deltacygtyd}) to an numerical integration of
equation~(\ref{eq:deltaceta}) and find the small correction
\begin{equation}
A_1(q) \rightarrow {A_1(q) \over 1 - 0.193 \sqrt{\fnu N_\nu} + 0.169 \fnu
N_\nu^{0.2} } \,,
\label{eqn:correction}
\end{equation}
for $\fnu \simlt 0.6$.
With the identities $\alpha_1 = p_{c}-1$ and $\tilde \alpha_1 = p_{cb}-1$, 
the derivation of the small-scale evolution [eq.~(\ref{eqn:deltacb})] is 
now complete.

\end{document}